\documentclass[12pt,a4paper]{article}

 \usepackage[dvips]{graphics}
 \usepackage{graphicx}
 \usepackage{afterpage}
 \usepackage{enumerate}
\begin{document}
 %%==========================================================

\title{\bf Two-dimensional MHD models of solar magnetogranulation.
Dynamics of magnetic elements}
 \author{\bf V.A. Sheminova}
 \date{}

 \maketitle
 \thanks{}
\begin{center}
{Main Astronomical Observatory, National Academy of Sciences of
Ukraine
\\ Zabolotnoho 27, 03689 Kyiv, Ukraine\\ E-mail: shem@mao.kiev.ua}
\end{center}

 \begin{abstract}
We present the results of a statistical analysis of the Doppler shifts and
the asymmetry  parameters of $V$ profiles of the Fe~I 
630.25~nm line produced by 2D MHD simulations  of
solar granulation. The realism of the  simulations 
tested using the magnetic ratio of Fe~I~524.71 and 525.02~nm lines. 
The Stokes spectra were synthesized in snapshots with a mixed polarity 
field having a mean magnetic flux density of 0.2~mT 
and mean unsigned field strength of 35~mT.
We found that downflows with a velocity of 0.5~km~s$^{-1}$ predominate, on the
average, in areas with some network magnetic elements at the disk center. 
 In separate  strong  fluxtubes average velocity is equal to 3~km~s$^{-1}$ and
the maximum velocity is 9~km~s$^{-1}$. In weak
diffuse magnetic fields upflows dominate. Their average  velocity is 
 0.5~km~s$^{-1}$ and maximal one is
3~km~s$^{-1}$. The  $V$-profile asymmetry depends on the spatial resolution.
The $V$ profiles synthesized with
high spatial resolution (35~km) have average amplitude and area
asymmetries $-1$\%, 1\%, respectively. The asymmetry scatter
is $\pm70$\%  for weak profiles and $\pm10$\% for
strong ones. The profiles with low spatial resolution (700~km)  have
average amplitude and area asymmetries $3$\%, -2\%, respectively.
Low spatial resolution is a reason why the  amplitude
asymmetry is always  positive and  greater than the area asymmetry 
in $V$ profiles observed. We found weak
correlation between the asymmetry of $V$ profiles and velocity.
Upflows cause negative asymmetry, on the average, and downflows
cause positive asymmetry. We examined  center-to-limb variations 
of vertical velocity in magnetic elements. Beginning from $\cos \theta
=0.9$, the average velocity abruptly increases from  0.5 
 to 2~km~s$^{-1}$
and then slightly varies closer to the limb. We found
nonlinear oscillations of vertical velocity  with power
peaks in the 5-minute and 3-minute bands. This
nonlinearity is caused by magnetic field strength fluctuations in fluxtubes.
The Doppler shifts and asymmetry parameters obtained for
space-average $V$ profiles are consistent with results of FTS
observations as well as with other observations made with higher
spatial resolution.
\end{abstract}
%-------------------------------------------------

\section{Introduction}
     \label{S-Introduction}
Magnetic fields on the Sun exhibit fine structure which is
conditioned by the fine structure of the solar surface in
general and its active regions in particular \cite{8}. The first
supporting evidence for the fine structure of magnetic fields
was found in the 1960s, initially in strong fields in spots and
later in quiet regions \cite{8}. Magnetic fields outside spots
were found to be unevenly distributed over the surface, in the
form of dense magnetic fluxes in very small areas ($\approx
1^{\prime\prime}$). Kilogauss strengths of these compact
magnetic features were initially measured in faculae and in the
supergranular network by the line-ratio method  \cite{24,45,47}.
These results were confirmed more than once in later studies.
According to \cite{45}, the typical size of magnetic
concentrations is 100--300~km, typical strength is 100--200~mT,
the velocity does not exceed 1.2~km~s$^{-1}$ and is 0.5
km~s$^{-1}$ on the average. The field strength scatter is very
small. The authors of  \cite{24} concluded that 90 percent of
the photospheric magnetic field is concentrated in compact
kilogauss structures and 10 percent is found in quiet regions
with a mean flux density of 0.2--0.3~mT. Further investigations
revealed that fine compact magnetic features have a filamentary
structure and look like tubes with their diameter growing with
height in the atmosphere (the mushroom or canopy effect). They
are tilted most probably at an angle of $10^{\circ}$
\cite{20,32}. The temperature in them is higher in the middle
and upper photosphere and is lower in the lower photosphere as
compared to the quiet Sun, and the chromospheric temperature
growth begins in deeper layers as compared to the quiet Sun
\cite{39}. Analyses of all available data revealed a wonderful
property of these magnetic structures --- despite some
distinctions in size, field strength, and orientation, they all
are much alike in form and in their magnetic and temperature
properties. Because of this, they were pooled to form a separate
class of solar magnetic structures and were called the
small-scale magnetic elements or magnetic flux tubes
\cite{36,39}. The number of flux tubes on the solar surface
increases with decreasing tube size. Although each magnetic
element adds very little to the general solar radiation flux,
millions of magnetic elements not only compensate for the energy
blocked in spots but even produce some additional radiation
(about 0.1 percent  \cite{40}). Magnetic flux tubes can form
clusters of various sizes. These clusters are assumed to form
larger magnetic structures such as the active supergranular
network, faculae, and plages. At present the smallest (120~km)
magnetic elements  resolved in observations are the so-called
bright points of magnetic network  \cite{16}.

Concurrent with the refinement of observation techniques,
theoretical models of magnetoconvection in quiet regions on the
Sun have been developed. Much research was devoted to numerical
simulation of magnetic elements, their formation and interaction
with convective motions (see review \cite{36}).

In this study we use two-dimensional MHD models developed by
Gadun \cite{4,18,33}. Unlike other researchers, he used an
original approach to the solution of the equations of radiation
magnetohydrodynamics \cite{5} which describe a compressible,
gravitationally stratified turbulent medium. The magnetic field
is described by the vector potential, so that the divergence of
magnetic field strength is always zero in the simulation region.
The initial magnetic field is specified by a bipolar magnetic
configuration. The first sequence of 2D MHD models \cite{4} of
completely nonstationary solar magnetogranulation extended over
a long time period (two hours of solar time). With such models,
the evolution of solar magnetic elements, their structure and
dynamics could be studied in detail on scales much smaller than
the spatial resolution of the present-day observations.
Three-dimensional models (e.g., \cite{42}) are certain to give
more realistic flow patterns, while two-dimensional models more
adequately represent small-scale phenomena, and they still
remain useful in studying the properties of magnetic elements.
Analyses of 2D simulations of convection \cite{17,35} and
magnetoconvection \cite{1,2,22,44} suggest that 2D models
reproduce quite well many features of 3-D convection. The major
results of 2D magnetoconvection simulations on granulation
scales obtained by Gadun were presented in \cite{4}.
Investigations into the mechanism of formation and destruction
of flux tubes and into their stability conditions revealed that
thermal flows are the most important factor in the evolution of
small-scale magnetic fields. Fragmentation of large-scale
thermal flows can result in the formation of vertical magnetic
elements from horizontal magnetic surface flows. This previously
unknown mechanism of flux tube formation was called the surface
mechanism \cite{5,18}. The mechanism of formation of kilogauss
flux tubes through convection collapse was also investigated,
and some spectropolarimetric manifestations of this process were
found \cite{14}. MHD models \cite{4} pointed to the emergence,
reconnection, and recycling of magnetic flux near the surface,
and these processes were demonstrated in \cite{34}.
Nonstationary MHD models \cite{4} can serve as an analog of some
observed photospheric regions with average unsigned strength of
magnetic fields of 40--50~mT. These heterogeneous models, which
are more realistic than two-component models, were also used to
calculate the Stokes profiles of photospheric lines
\cite{11,13,34}. In particular, the analysis of the $V$ profiles
of the IR line Fe~I $\lambda$ 1564.8~nm \cite{13} gave a
magnetic field intensity distribution which is in good agreement
with observations of magnetic fields in quiet regions \cite{27}
and which confirms the two-component structure of intra-network
fields.

The magnetic field in the solar plasma is closely associated
with the velocity field, and this association inside sunspots as
well as in the quiet solar atmosphere still is one of the most
difficult problems in solar physics. The freezing of magnetic
field in the moving plasma (or its rigid connection with
velocity field) can be weakened due to the fine structure of
this magnetic field \cite{8}. That is why the study of dynamic
processes in magnetic elements, and inside strong flux tubes in
particular, by the magnetoconvection simulation techniques is of
current interest, inasmuch as the magnetic elements still are
not spatially resolved because of their small horizontal size.
Numerical MHD simulations \cite{1,44} suggested that a large
variety of dynamic processes should be expected on scales of
several hundred kilometers inside and outside flux tubes. Recent
highly accurate polarimetric observations in the line Fe~I
$\lambda$ 630.25~nm with a spatial resolution better than 700~km
\cite{37} yielded some statistical relations between the
observed Stokes $V$-profile parameters which are indicative of
substantial systematic motions in the magnetic elements in the
network, faculae, and quiet regions. We believe that the
reproduction of some observed spectropolarimetric effects
\cite{28,32,37} with the use of 2D~MHD models \cite{4} is of
obvious interest.

The main purpose of this study is to determine the dynamical
characteristics of magnetic elements in the solar photosphere.
Below we give some principal results of observations of the
magnetic element dynamics, a description of the 2D MHD models we
use, and the results of the application of the line-ratio method
to these models. We also analyze our calculations of the Stokes
$V$ profiles, their shifts, asymmetries, and center-to-limb
variations on scales smaller than or comparable to the spatial
resolution of present-day observations.
%-------------------------------------------------

\section{Observed shifts and asymmetries of V profiles }
     \label{ Observation }

Studies based on complex magnetographic observations of magnetic
fields and line-of-sight velocities revealed that the vertical
velocities are equal to zero, on the average, at the
chromospheric network boundaries as well as inside network cells
\cite{9}, while the line-of-sight velocities determined from
spectropolarimetric observations were contradictory \cite{36}.
Early extensive polarimetric observations with Fourier
spectrometers (FTS) with high spectral resolution and low
spatial and time resolutions (about $10^{\prime\prime}$ and 30
min) detected no significant systematic motions (above 0.25
km~s$^{-1}$) inside flux tubes \cite{41}. This result differed
essentially from a mean velocity of 0.5~km~s$^{-1}$ derived
earlier in \cite{45}. Only observations with high spatial
resolution ($\approx1^{\prime\prime}$) {32} detected a small
predominance of downflows with a mean velocity of 0.2
km~s$^{-1}$ and the dependence of this velocity on the size of
magnetic elements. In regions with small filling factor,
downflows were stronger and the velocity scatter increased from
$-0.5$~km~s$^{-1}$ to 1.5~km~s$^{-1}$. The mean velocity
measured in \cite{20} was about 0.8~km~s$^{-1}$, and the scatter
increased with decreasing amount of polarization. Theoretical
simulations of the interaction between convection and magnetic
field \cite{1,22} also confirmed the existence of systematic
motions outside and inside flux tubes on small spatial and
temporal scales. A most impressive evidence that dynamic
processes are stronger inside magnetic elements was given in
recent studies \cite{27,37}. Systematic downward motions
observed with a spatial resolution of 0.8--$1^{\prime\prime}$
were found to have mean velocities of 0.5--0.7~km~s$^{-1}$. In
magnetic elements with small filling factor the velocities were
as high as $\pm5$~km~s$^{-1}$, while they were smaller in
elements with large filling factor or in clusters. Some evidence
for large horizontal velocities of about 2~km~s$^{-1}$ in
magnetic elements was given in \cite{51}.

In addition to the zero-crossing Doppler shifts, the asymmetry
of observed $V$ profiles is used for the diagnostics of motions
in flux tubes. This asymmetry detected for the first time in
\cite{38} from FTS observations near the disk center was
``blue'', i.e., the amplitude and area of the shortwave peak
were greater, on the average, than in the longwave peak, the
amplitude asymmetry being several fold greater than the area
asymmetry. The general trends found in \cite{38} have been
confirmed by observations made with various resolutions, but the
asymmetry magnitude slightly decreased with growing resolution.
Numerous investigations demonstrated that the $V$ profiles owe
their asymmetry to the combined effect of the magnetic field and
velocity field gradients inside flux tubes and around them
\cite{36}. A large number of profiles with extremely large
asymmetries were recently found in observations with resolutions
below 700~km. For example, three percent of all Fe~I $\lambda$
630.25~nm line profiles observed in the network regions and in
faculae have such asymmetries \cite{37}. Atypical $V$ profiles,
especially those which have one wing only, are of special
interest. They were shown in MHD simulations to appear mainly at
the periphery of magnetic elements \cite{34}. Extremely large
asymmetry can be caused by strong downflows below the canopy
near flux tube boundaries.
%-------------------------------------------------
%%%%%%%%%%%%%%%%%%%%%%%%%%%%%%%%%%%%%%%%%%% Figure 1
  \begin{figure}
\centerline{
\includegraphics [scale=0.9]{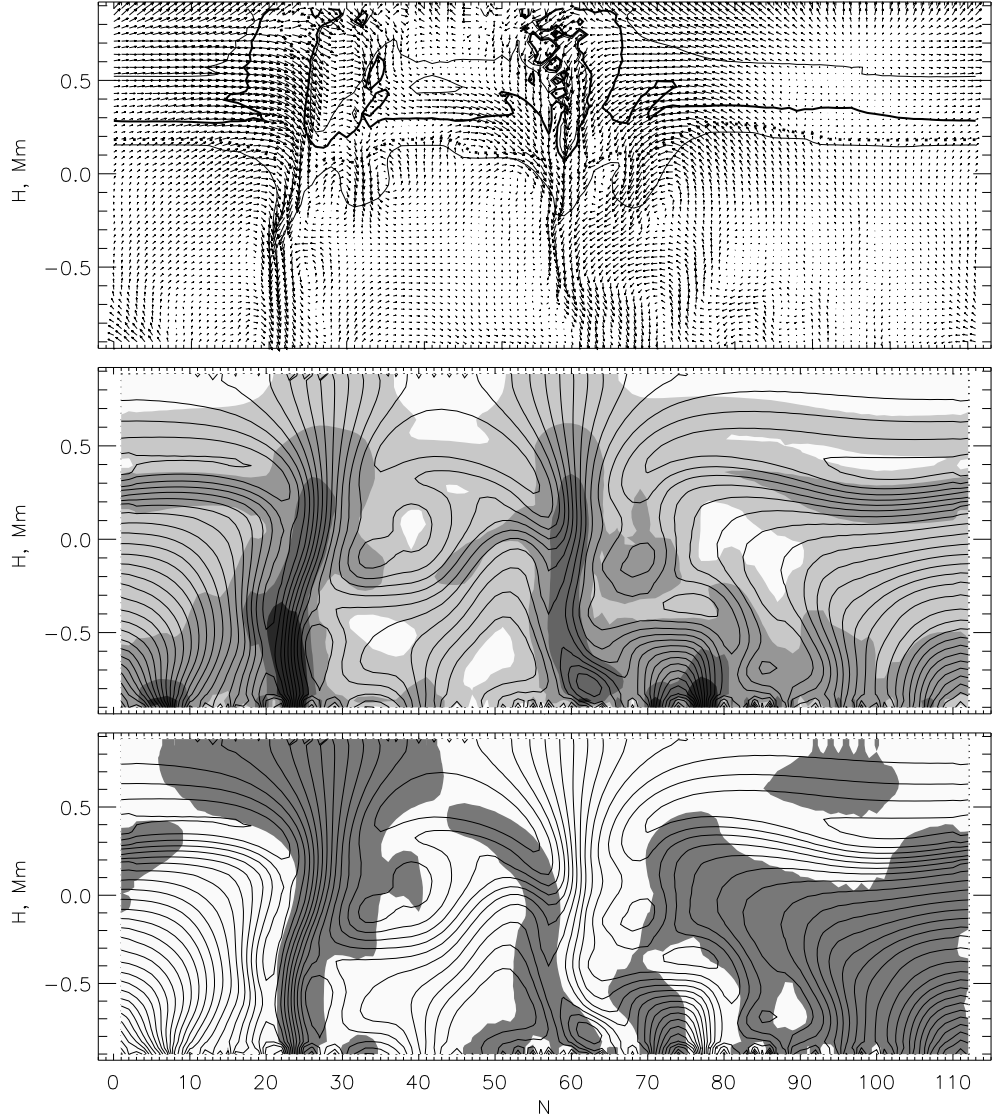}}
% \hfill
%
\vspace{0.1cm}
 \caption
{A snapshot from 2D MHD simulation \cite{4} at moment 95.5 min:
a) velocity field (arrow length is proportional to velocity
magnitude) and isotherms (from top to botton) corresponding to
4000 K (thin line), 5000 K (thick line), 6500 K (dotted line),
and 10000 K (thin line); b) field lines and magnetic field
strength (shading density is proportional to field strengths of
1, 20, 50, 100, 150~mT); c) field lines and field polarity (grey
area shows positive polarity and white area negative polarity).
Vertical axis) geometric height $H$, horizontal axis) column
number $N$ in the MHD model; distance between the columns is 35
km.} \label{F-1}
 \end{figure}
%%%%%%%%%%%%%%%%%%%%%%%%%%%%%%%%%%%%%%%%%%%%%%%%%%%%%%%%%%%%

So, polarimetric observations with high spatial resolutions
demonstrated that the dynamical characteristics of magnetic
elements have a much wider range than that assumed before. What
can we expect from observations in which the structure of
magnetic elements will be resolved? We tackled this question
with the use of the results of the 2D MHD simulation of magnetic
elements and the synthesis of the Stokes profiles of the Fe~I
$\lambda$ 630.25~nm line.

\section{Two-dimensional MHD models}

The sequence of MHD models we use here was described in detail
in \cite{4,5,18}. The simulated region 3920$\times$1820~km in
size had 112 vertical columns (rays) with a 35-km spatial step.
The starting  magnetic field  had  bipolar decreasing with
growing height, the mean unsigned field strength was 5.4~mT. The
magnetic field was evolving in the course of two-hour
simulation, and its unsigned strength grew to 50~mT on the
average. Strong flux tubes are formed after a lapse of 50 min of
simulation. They disintegrate and form again in the simulation
region. In one of the snapshots displayed in Fig.~1 two
kilogauss flux tubes of different polarities can be seen to
approach each other and to destroy one the other a short time
later. The horizontal size of flux tubes at the level of visible
surface ($\log \tau = 0$) varies from 35~km to 350~km in the
course of evolution, and the maximum strength of their fields
varies correspondingly from 40~mT to 250~mT. The largest area
(about 50 percent) is occupied throughout the simulation time by
flux tubes with diameters of 80--180~km and field strengths of
100--200~mT \cite{4}.

Our study is based on a sequence of the snapshots of half an
hour's duration after 94 minutes of simulation with a 30-s step.
This is a period when strong flux tubes are intensely evolving.
Two flux tubes of opposite polarities are often formed in the
simulation region. As a rule, flux tubes are accompanied by
nearby weak fields of opposite polarity. In upper lateral layers
of flux tubes hot regions appear as a result of deceleration of
strong granulation flows near the tube walls. These hot regions
may be analogs of observed bright points \cite{16}.
Predominantly horizontal magnetic fields are observed in the
central parts of granular cells. All these peculiarities of the
simulated magnetogranulation can be seen in the vertical section
of the simulation region (Fig.~1). We assume that the selected
snapshots sequence represents the quiet Sun regions with some
network elements because the unsigned field strength
distribution mode is 35~mT in the simulation region and the flux
density is 0.2~mT \cite{13}.

\section{Testing the MHD models}
     \label{Testing }

The MHD models we use were tested in \cite{11}. Here we continue
to test them by the method of field strength ratio measured in
two spectral lines \cite{45}. This well-known method was widely
used for the diagnostics of strong magnetic fields outside
active regions on the Sun in low spatial resolution spectra
(1--$4^{\prime\prime}$) \cite{19,24,25,47}, and now it is
applied to high-resolution spectra ($< 1^{\prime\prime}$)
\cite{25,52}. The effectivity of the line-ratio method depends
on the model atmosphere chosen for the field strength
calibration, so that the method also needs testing with models
which are more realistic than the two-component models commonly
used for calibration \cite{25}. Applying this method to MHD
models, we can test the models and the method itself.

As in observations, we chose two Fe~I lines -- $\lambda\lambda$
524.71 and 525.02~nm. Their atomic parameters are nearly the
same, and only the Lande  factors differ (their ratio is 1.5).
Therefore, the ratios of calibrated magnetograph signals in two
lines -- $B_{525.0}{/}B_{524.7}$ or $V_{525.0}{/}(1.5 V_
{524.7})$ -- should be insensitive to all atmospheric parameters
except field strength, magnetic field inclination, and velocity
field. As shown in \cite{7,46,47}, the ``magnetic ratio'' also
depends on the position of the selected $V$-profile section with
respect to the profile center -- as the magnetograph slit
approaches the line center, the ratio decreases. The line ratio
also depends on the filling factor in the resolved area and on
lateral magnetic field profile. Thus, the magnetic field
strength found by the line-ratio method depends on the spatial
and spectral resolutions of observed spectra.
%%%%%%%%%%%%%%%%%%%%%%%%%%%%%%%%%%%%%%%%%%% Figure 2
  \begin{figure}
%\centerline{
\includegraphics [scale=0.5]{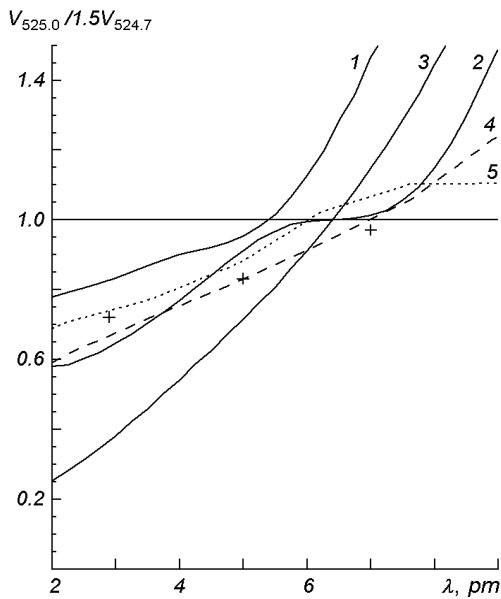}
%}
 \hfill
\vspace{0.1cm}
\parbox[b]{7.5cm}{\vspace{0.1cm}
 \caption
 {Ratio of $V$ profiles of Fe~I lines $\lambda\lambda$
525.0 and 524.7~nm as a function of the distance from
$V$-profile center. Solid curves were obtained from MHD models:
1) for region with one subkilogauss flux tube, 2) for region
with two kilogauss flux tubes, 3) for axis of kilogauss flux
tube; plusses) observations \cite{7} outside active regions;
curve 4) observations \cite{47} for strong faculae; curve 5)
observations \cite{47} for weak faculae.} \label{F-2}
}
 \end{figure}
%%%%%%%%%%%%%%%%%%%%%%%%%%%%%%%%%%%%%%%%%%%%%%%%%%%%%%%%%%%%

We determined the ratio of $V$ signals as a function of distance
to line center $MLR ( \Delta \lambda )=V_{525.0}( \Delta \lambda
)/[1.5 V_{524.7}( \Delta \lambda ) ]$ for the profiles in
different simulation regions. The Doppler shifts of $V$ profiles
with respect to the central wavelength of $I$~profiles were
compensated. Figure 2 illustrates three $MLR ( \Delta \lambda )$
relations. One relation was derived from profiles averaged over
a $3^{\prime\prime}$ region with a relatively weak subkilogauss
flux tube ($B_0\leq 90$~mT), the second relation refers to an
area of the same size with two strong flux tubes with $B_0\leq
190$~mT, and the third relation is based on the profiles
corresponding to the axis of a strong flux tube with
$B_0\geq180$~mT at the level $\log \tau_5=0$ (without spatial
averaging). We also plotted the relations for the ratios $B_
{525.0}{/}B_{524.7}$ (observations from \cite{7}) and
$V_{525.0}( \Delta \lambda){/}[1.5 V_{524.7}(\Delta \lambda )]$
(FTS observations of regions with strong and weak faculae
\cite{48}). The magnetic ratio is greater at greater distances
from line center and is greater than unity in distant line
wings. At lower resolutions, the relation is less steep and the
magnetic ratio is smaller (cf curves 2 and 3). The run of
relation 2 (the region with two strong flux tubes and with low
resolution) is in good agreement with observation data, and this
supports our earlier inference \cite{11} that the theoretical
MHD models \cite{4} represent quite well the actual magnetic
fields on the Sun.

The run of the $MLR ( \Delta \lambda )$ relations derived for
different magnetic fluxes from MHD models is the same as the run
derived earlier with the use of two-component models
\cite{7,46}. These relations roughly represent various cases of
magnetic saturation and various long-wave shifts of $V$ signals
in different sections of the profiles of two lines. Kilogauss
fields not only diminish the $V$-profile amplitude but
substantially change the profile shape as well. In strong fields
the $V_{525.0}$ profile is lower and wider than the $V_{524.7}$
profile, and the ratio of magnetic saturations in two lines
becomes reverse at a distance of 5--7 pm from the line centers:
$MLR$ becomes greater than unity. So, the $MLR ( \Delta
\lambda)$ relations we derived clearly demonstrate that the
magnetic fields measured by the line-ratio method heavily depend
on the line section ($\Delta \lambda$) chosen for the
measurements.

The ratio of $V$-profile amplitudes $MLR =
a_{V,525.0}{/}(1.5a_{V,524.7})$, which is independent of $\Delta
\lambda$, is sometimes used in polarimetric observations
\cite{47}. We determined the amplitude ratio for 5824 $V$
profiles calculated in every MHD model column (Fig.~3a). There
is a correlation between the ``magnetic ratio'' and the
amplitude $a_{V,525.0}$: the ratio decreases, on the average,
with increasing $V$-profile amplitude, and this suggest that the
magnetic field increases. We also determined $MLR$ for profiles
averaged with a $1^{\prime\prime}$ resolution. This spatial
averaging markedly affected the $V$-profile amplitudes and
consequently the magnetic ratio (Fig.~3b ); the $MLR$ scatter
increased in every interval and the correlation between $MLR$
and profile amplitude deteriorated. The number of atypically
shaped profiles (with one wing or with several wings) was found
to grow among weaker $V$ profiles, and this can affect the
accuracy with which mean $V$-profile amplitudes are calculated.
It should be noted that the Fe~I lines $\lambda\lambda$ 524.71
and 525.02~nm have very narrow profiles which are highly
sensitive to temperature, so that the effect of extreme
asymmetry of weak $V$ profiles is stronger for these lines, and
this accounts for the greater scatter in $MLR$ for small
$V$-profile amplitudes.

We calculated the $MLR$ values in order to transform the
magnetic ratios into field strengths. This was done with the
data for the MHD model shown in Fig.~1. We calculated $MLR$ and
effective heights of $V$-profile peak formation for each model
column. Having found the level at which the method locates the
magnetic field, we determined the field strength for this level
directly from the model and plotted the $MLR$--$B$ relation
(Fig.~3c). We found from this relation that a decrease of $MLR$
from 0.94 to 0.82 (Fig.~3a) corresponds to a growth of $B$ from
30 to 62~mT (Fig.~3c). This means that, according to the
%%%%%%%%%%%%%%%%%%%%%%%%%%%%%%%%%%%%%%%%%%% Figure 3
  \begin{figure}
%\centerline{
\includegraphics [scale=0.45]{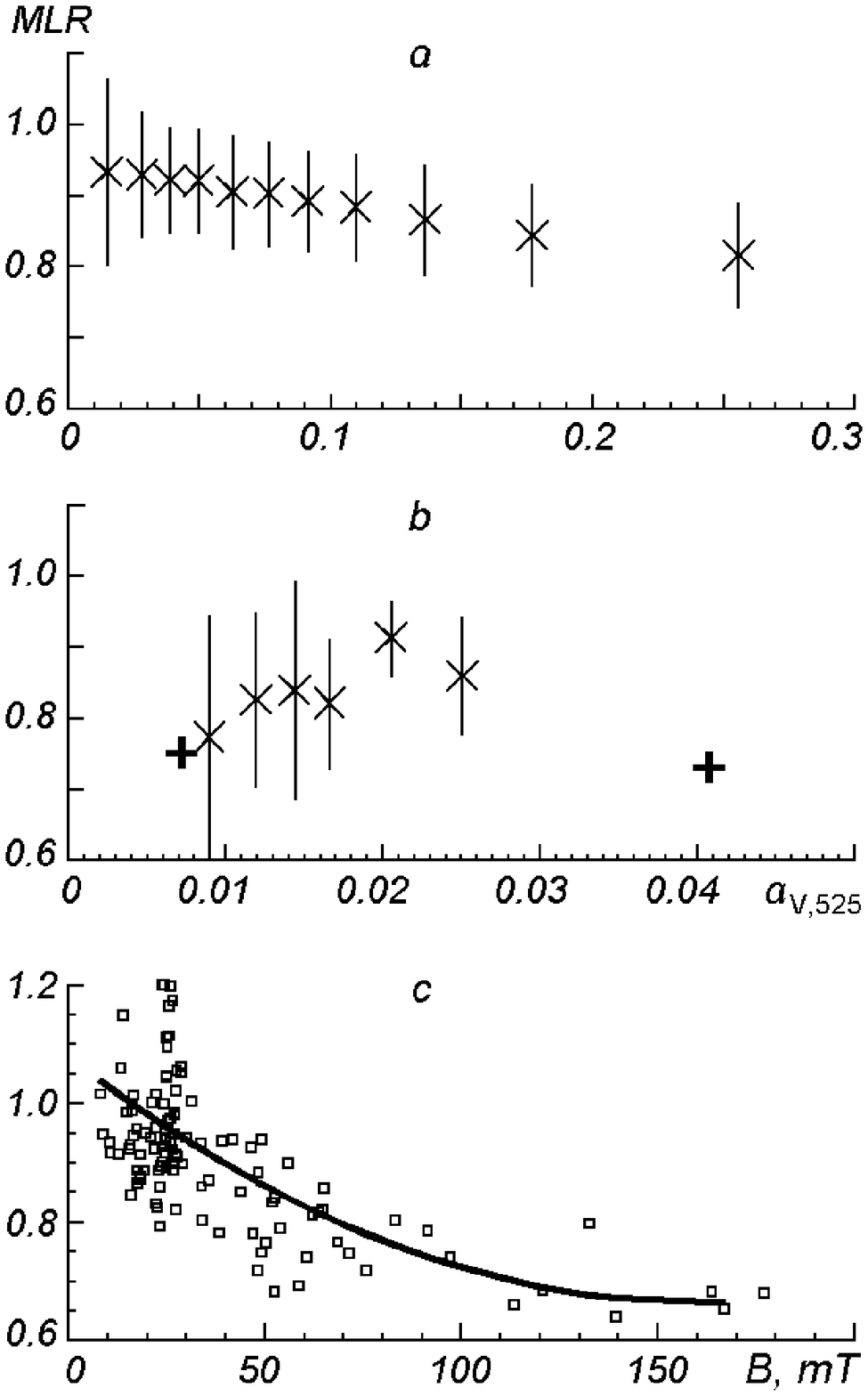}
%}
 \hfill
\vspace{0.1cm}
\parbox[b]{7.5cm}{\vspace{0.1cm}
 \caption
{Magnetic ratio $MLR =a_{V,525.0}{/} ( 1.5a_{V,524.7})$: a)
without averaging, b) with spatial averaging over 700~km.
Standard deviations are shown for some intervals with the same
number of points; c) dependence of MLR on magnetic field
strength B derived from the data of the snapshot shown in
Fig.~1.} \label{F-3}}
 \end{figure}
%%%%%%%%%%%%%%%%%%%%%%%%%%%%%%%%%%%%%%%%%%%%%%%%%%%%%%%%%%%%
line-ratio method, the field strength in the simulation region
can be $60\pm 30$~mT at the most, i.e., the method did not fix
any strong fields about 150~mT appropriate for the MHD models we
use. The averaged magnetic line ratio is $0.88\pm 0.10$, and the
corresponding mean field strength in the simulation region is 43
mT, in good agreement with the simulation results for our MHD
model sequence (40--50~mT) and with the results obtained in
\cite{52} by the line-ratio method applied to the $V$ profiles
of the same two lines observed in quiet regions with high
spatial resolution. No significant deviations of $MLR$ from
unity were found in \cite{52}, and the intra-network fields were
concluded to have no kilogauss structures; the typical field
strength was estimated at 20--50~mT. The coincidence of the data
from \cite{52} with the mean field strength in the MHD~models we
used allows us to suggest that our simulation region is similar
to quiet solar regions.

Thus, the testing of MHD models by the line-ratio method with
the use of the Fe~I lines $\lambda \lambda$~524.71 and 525.02~nm
showed a satisfactory agreement between the calculated and
observed magnetic ratios as functions of the distance from the
$V$-profile center. This suggests that the MHD models \cite{4}
describe adequately the fine structure of actual magnetic fields
of the quiet Sun. The testing also demonstrated that fields
stronger than 100~mT with filling factors of 1--5 percent cannot
be detected by the line-ratio method in quiet regions on the
Sun, since this method depends on spatial resolution.
Nevertheless, the method can give a reliable estimate for the
mean field strength in quiet regions.

\section{Results of analysis of synthesized V profiles}
     \label{Results}

The Stokes profiles of the Fe~I line $\lambda$ 630.25~nm were
calculated for every column in the MHD models by integrating the
Unno-Rachkovskii equations for the polarized radiation transfer
in the LTE approximation \cite{10}. We chose this moderately
strong photospheric line for the $V$-profile analysis because it
is most often used, together with the Fe~I line $\lambda$ 630.15
nm, in polarimetric measurements \cite{15,28,30,31,37}. A
telluric O$_2$ line which is observed in the $I$-profile wing
does not affect the $V$ profile. The line $\lambda$ 630.25~nm is
preferred to the magnetic Fe~I lines $\lambda\lambda$ 524.7 and
520.5~nm not only because of its unblended $V$ profile but also
due to its high sensitivity to magnetic field and much smaller
sensitivity to temperature irregularities. The $V$ profile of
Fe~I $\lambda$ 630.25~nm is formed rather deep in the
photosphere, at the level $\log \tau_5 = -1$, on the average,
and thus it can be calculated in the LTE approximation. It
should be stressed that the dynamical characteristics of
magnetic elements studied with the use of this line refer to the
same photospheric level. This is particularly true for the
velocity and magnetic field gradients which can suddenly change
in highly inhomogeneous media, for example, at the periphery of
compact magnetic features. That is why the asymmetry magnitude
and sign, which attest to the existence of gradients, as well as
the relationship between the asymmetry and other parameters
found in our study can differ from the results obtained with the
use of other lines.

We compared our calculations to the FTS observation data kindly
made available by J.~Stenflo and S.~Solanki. These observations
were made in quiet network elements and in active facula regions
at the McMath telescope in 1979; the Fourier spectrograph had a
high spectral resolution (420 000) and low spatial and time
resolutions ($10^{\prime\prime}$ and 35--52 min, respectively)
\cite{48}. We used only the $V$ profiles of Fe~I and Fe~II lines
in the wavelength range from 445.0 to 557.0~nm. The observed $V$
profile parameters were calculated by Solanki's codes, and the
absolute zero-crossing shifts were determined by the method
proposed in \cite{3}.

We analyzed the following $V$-profile parameters. 1. Mean
amplitude of blue ($b$) and red ($r$) wings $a_V=(| a_b |+|a_r
|)/2$. 2. Doppler zero-crossing shift with respect to the
absolute wavelength (transformed into the line-of-sight velocity
$V_z$ by the standard formula). 3. Amplitude asymmetry $\delta
a=(| a_b |-| a_r |){/}(| a_b |+|a_r |)$ and area asymmetry
$\delta A=| A_b |-| A_r |){/}(| A_b |+| A_r |)$. The blue and
red wing amplitudes ($a_b$ and $a_r$) and the zero-crossing
position were calculated by fitting the corresponding profile
sections to polynomials. The areas $A_b$ and $A_r$ were obtained
by integrating the $V/I_c$ profile from zero-crossing to a
0.5-percent level in the red and blue wings.

By analogy with the observation data from \cite{37}, we used
only the profiles with amplitudes greater than 0.15 percent and
regularly shaped profiles, i.e., only those with two wings of
different signs and with one zero-crossing. The number of such
profiles was 3755; the rest were either very weak or abnormal in
shape (symmetric, with one wing only, or with several
zero-crossings), and they were excluded from the analysis.

{\bf Mean $V$-profile amplitudes}. Figure 4 (upper part)
demonstrates the  $a_V$ distribution obtained from $V$-profiles
calculated without spatial averaging and with a
$1^{\prime\prime}$ averaging. When the histogram for all
profiles is compared to the histogram for regularly shaped
profiles (without strong anomalies), one can see that weak
profiles are less numerous in the second case due to highly
asymmetry of weak  profiles. Spatial averaging of profiles
essentially changes their distribution. Recall that the maximum
diameter of simulated flux tubes is about 350~km, so that the
profiles averaged over a 700-km area do not resolve the tube
structure, and profiles with amplitudes above 10 percent are
absent in the histogram.
%%%%%%%%%%%%%%%%%%%%%%%%%%%%%%%%%%%%%%%%%%% Figure 4
  \begin{figure}
\centerline{\hspace{1.5cm}
\includegraphics [scale=0.8]{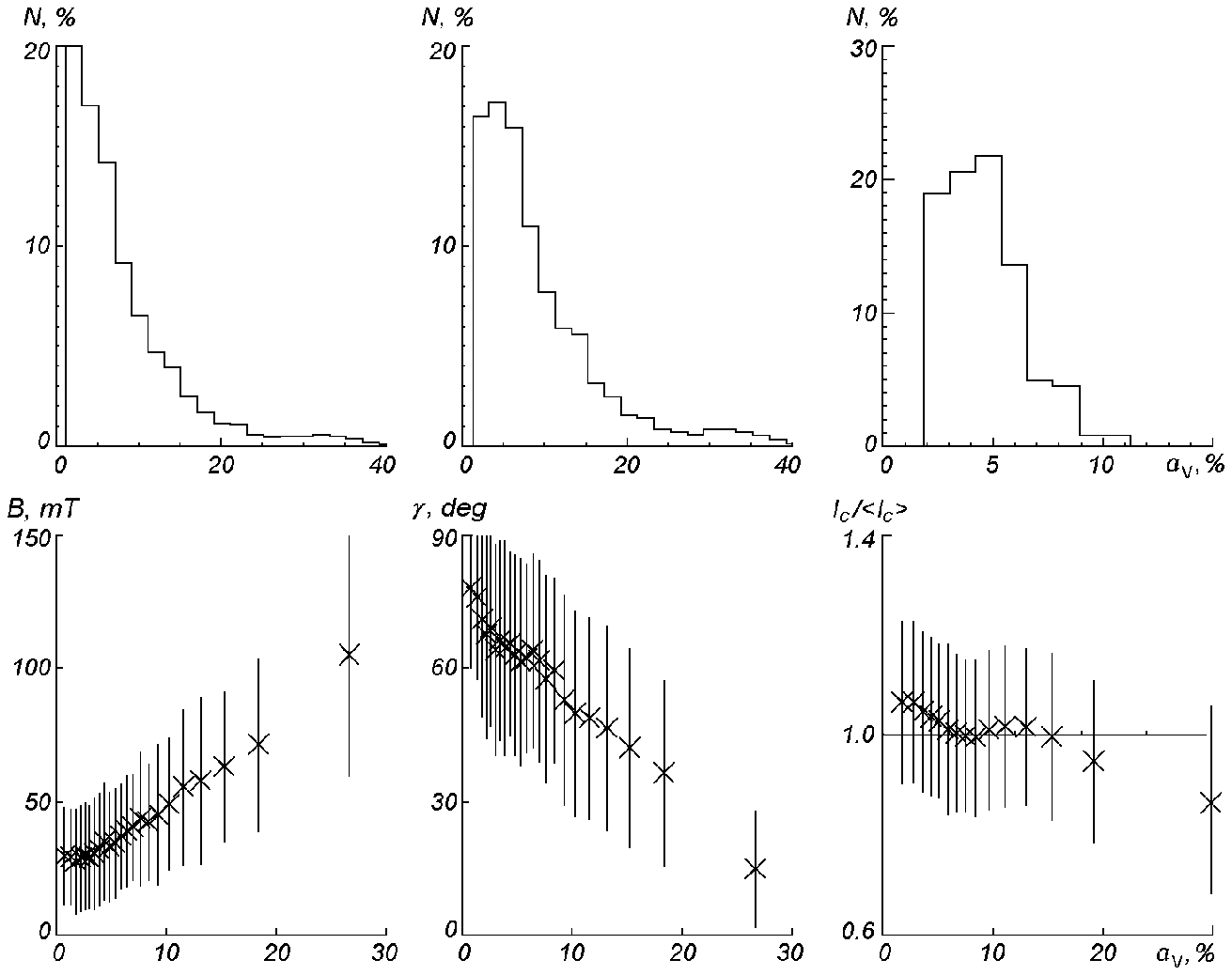}}
 \hfill
\vspace{0.1cm}
 \caption
{ Upper plots) distributions of $V$ profile amplitudes $a_V$
from 52 MHD snapshots: at the left) for all 5824 profiles ($
\overline{a_V} = 7$\%); at the center) with abnormal profiles
excluded (3755 profiles, $\overline{a}_V = 9$\%); at the right)
for profiles spatially averaged with a 700-km step (243
profiles, $\overline{a_V} = 4$\%). Lower plots) unsigned field
strength, field inclination angle, and contrast of continuum
intensity as functions of amplitude $a_V$.} \label{F-4}
 \end{figure}
%%%%%%%%%%%%%%%%%%%%%%%%%%%%%%%%%%%%%%%%%%%%%%%%%%%%%%%%%%%%

In the analyses of observed Stokes profiles, $a_V$ is most often
used as a substitute for filling factor or as an indicator of
the longitudinal magnetic field. The statistical relations for
$a_V$ shown in Fig.~4 (lower part) were averaged over the
intervals of the field strength $B$ and field inclination
$\gamma$, they were obtained directly from the MHD models for
the optical depth $\log\tau_5 = -1$ and from the continuum
intensity contrast $I_c/<I_c>$. Here $I_c$ is the continuum
intensity calculated for a specific model column and $<I_c>$ is
the intensity averaged over the entire simulation region in the
model. There is a close correlation between $B$ and $a_V$ as
well as between $\gamma$ and $a_V$. The relationship between
$I_c/<I_c>$ and $a_V$ is more complicated --- the weakest
$V$-profiles correspond to high-contrast areas, the strongest
profiles correspond to low-contrast areas, while moderately
strong profiles represent areas with slightly changing contrast.

So, the nearly linear $a_{V}$--$B$ relation found suggests that
$a_V$  can be used as a magnetic field indicator in the analyses
of Stokes profiles.

{\bf Zero-crossing shifts.} The line-of-sight velocities ($V_z$)
of the magnetized plasmas flows  only are derived from the
Doppler shifts of $V$ profiles or the so-called zero-crossing
shifts.  Positive (red) shifts point to downflows and negative
(blue) shifts to upflows. Figure 5 (upper part) displays the
Doppler shifts  converted into $V_z$ as functions of three
principal parameters derived usually from the observed
$V$-profiles:  $a_V$ (indicator of magnetic field strength),
magnetic vector inclination $\gamma$, and intensity contrast
$I_c/<I_c>$.  The inclination angle was derived from the
amplitude ratio for Stokes profiles: $\tan^2 \gamma \approx Q^2
+ U^2 )^{1/2}/V^2$.
%%%%%%%%%%%%%%%%%%%%%%%%%%%%%%%%%%%%%%%%%%% Figure 5
  \begin{figure}
\centerline{\hspace{2.5cm}
\includegraphics [scale=0.9]{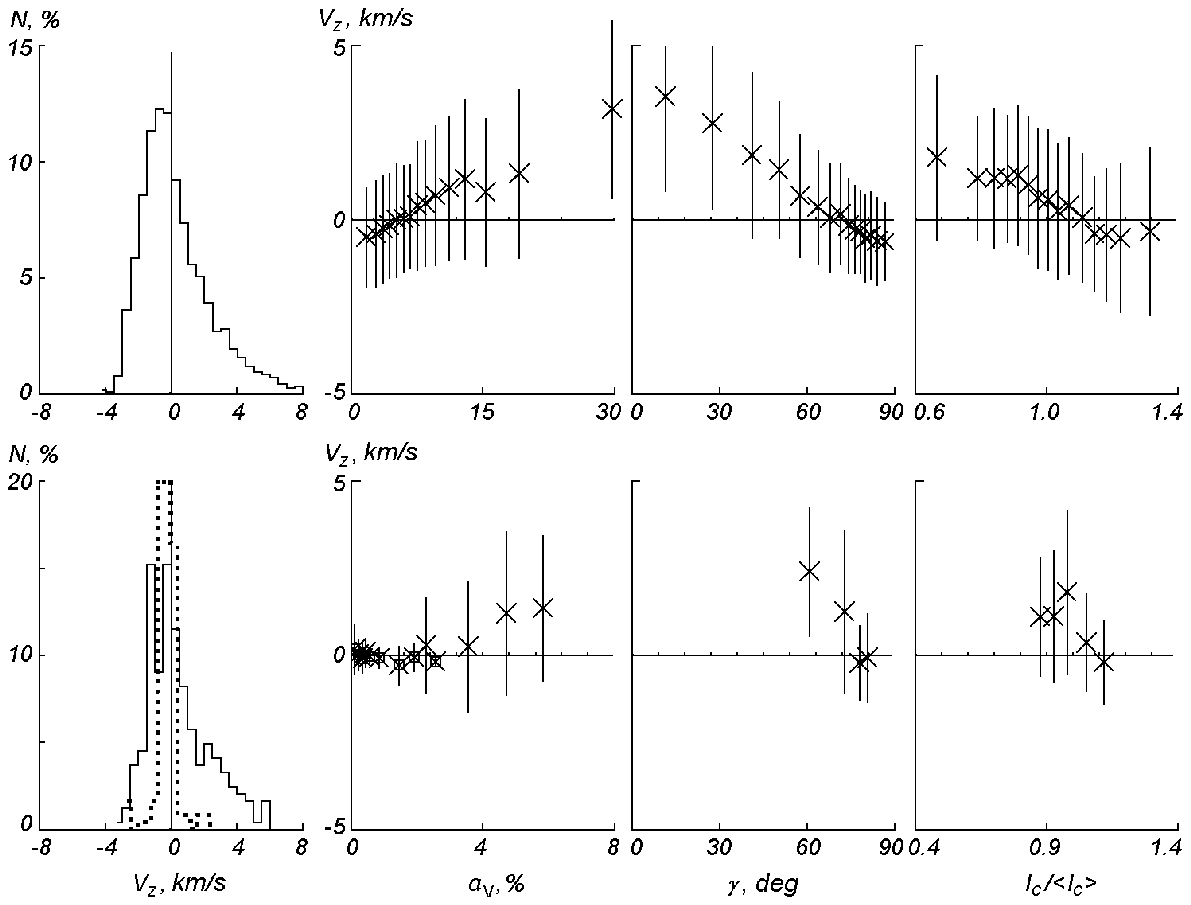}}
 %\hfill
%
\vspace{0.1cm}
 \caption
{$V_z$ histograms derived from the zero-crossing shifts of $V$
profiles, and the $V_z$ dependence on $a_V$,  $\gamma$, and
$I_c/<I_c>$: upper plots) all profiles, lower plots) profiles
averaged with a 700-km step. Dotted line and squares) FTS
observation data.} \label{F-5}
 \end{figure}
%%%%%%%%%%%%%%%%%%%%%%%%%%%%%%%%%%%%%%%%%%%%%%%%%%%%%%%%%%%%

The distribution of $V_z$s (Fig.~5, left) points to a
well-marked asymmetry between upflows and downflows -- a
fundamental property of magnetoconvection in a matter with
frozen magnetic field. The velocity distribution maximum lies at
$-1$~km~s$^{-1}$ approximately. A mean velocity of 0.53
km~s$^{-1}$ and a~wide velocity range from $-3$~km~s$^{-1}$ to 9
km~s$^{-1}$ suggest that downflows dominate in the magnetized
plasma. There is a correlation between velocity and principal
parameters of $V$ profiles . The downflow (positive) velocity
increases  to 3~km~s$^{-1}$, on the average, with increasing
$a_V$ ($B$), with decreasing $\gamma$ and $I_c/<I_c>$, and this
suggests that the most rapidly moving plasma resides in strong
(120~mT) nearly vertical ($10^\circ$) flux tubes with low
contrast (0.7), which corresponds to darker intergranular lanes.
The above correlations are in accord with the physical
properties of convective motions which affects the distribution
of LOS  velocities and magnetic fields in  photosphere (see
velocity field of the convective motions in the snapshot,
Fig.~1).

The spatial averaging of profiles decreases the fraction of very
weak profiles with negative velocities and very strong profiles
with high positive velocities (the lower part of Fig.~5). The
spatial averaging increased the mean value of $V_z$  to 0.72
km~s$^{-1}$. All relations in the lower part of  Fig.~5
preserved the trends pronounced in the upper plots.

Our results agree satisfactorily with the observations of
magnetic regions outside activity centers analyzed in \cite{37},
where the mean velocity was estimated at 0.73~km~s$^{-1}$ and
the velocities ranged from $-6$ to 6~km~s$^{-1}$. In \cite{27}
$\overline{V}_z$ was 0.25~km~s$^{-1}$ and the range was from
$-3$ to 5~km~s$^{-1}$. We also plotted the FTS observation data
from \cite{48} ($\overline{V}_z = -0.04$~km~s$^{-1}$) in Fig.~5.
Our results are in good agreement with observed velocities in
the regions of small $a_V$.
%%%%%%%%%%%%%%%%%%%%%%%%%%%%%%%%%%%%%%%%%%% Figure 6
  \begin{figure}
%\centerline{
\includegraphics [scale=0.7]{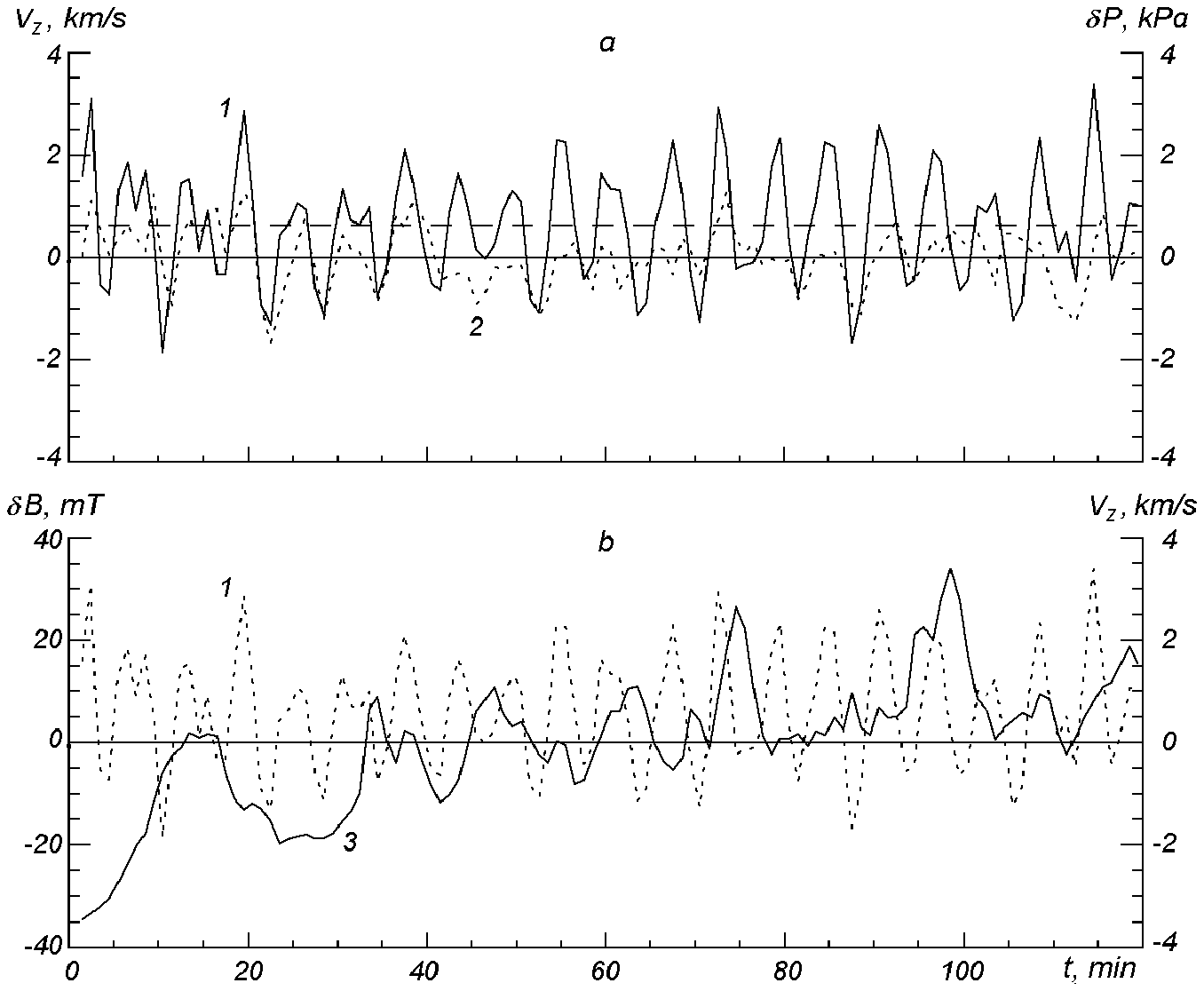}
%}
 \hfill
\vspace{0.1cm}
\parbox[b]{6.cm}{\vspace{0.1cm}
 \caption
{Oscillations derived from 2D MHD model of solar granulation. 
a)~Vertical velocity $V_z$ (curve 1), gas pressure fluctuations
$\delta P$ (curve 2).  Dashed straight line) mean velocity of
$0.6\pm 1.1$~km~s$^{-1}$. b)  $V_z$ (curve 1),
unsigned magnetic field strength $\delta B$  at the level $\log\tau_5 =
0$ (curve 3). $V_z$,  $\delta P$, and $\delta B$  are averaged
over whole simulation region.} \label{F-6}}
 \end{figure}
%%%%%%%%%%%%%%%%%%%%%%%%%%%%%%%%%%%%%%%%%%%%%%%%%%%%%%%%%%%%

{\bf Oscillations.} Although wave processes in spatially
unresolved flux tubes are difficult to detect directly,
oscillatory motions with a 5-minute period along flux tubes have
been reported \cite{39}. There is also some evidence for
oscillations with a shorter period \cite{51}. Observations with
high spatial resolutions \cite{15} revealed velocity
oscillations with an amplitude of about 1~km~s$^{-1}$. Theory
predicts a wide variety of wave modes in solar flux tubes and
their importance in the heating of the chromosphere and corona.
According to the 2D MHD simulation \cite{4} velocities of
vertical flows in magnetized plasmas can vary in a wide range
(see Fig.~5). It points to an oscillatory instability of
magnetogranulation regions. Here we examined the temporal
changes of velocity $V_z$ derived from shifts of $V$ profiles
averaged over in the whole simulation region for the complete
two-hour sequence of MHD models with a~1-minute interval.
Fig.~6a shows obtained  vertical velocities as functions of
time. We also calculated the power spectrum of velocity
oscillations (Fig.~7a), where two power peaks can be clearly
seen in the bands at 5~min ($2<\nu<4.5$~mHz) and 3 min ($4.5<
\nu < 7$~mHz). The power maximum lies at $\nu \approx 2.8$ mHz
(5.9 min), and a smaller peak lies at $\nu \approx 4.7$~mHz (3.5
min). A 99-percent confidence level shown by a horizontal line
in the figure was calculated in accordance with \cite{23} from
the mean power in the band 5.5--8.3 mHz. The Nyquist frequency
is equal to 8.3 mHz.
%%%%%%%%%%%%%%%%%%%%%%%%%%%%%%%%%%%%%%%%%%% Figure 7
  \begin{figure}
%\centerline{
\includegraphics [scale=0.55]{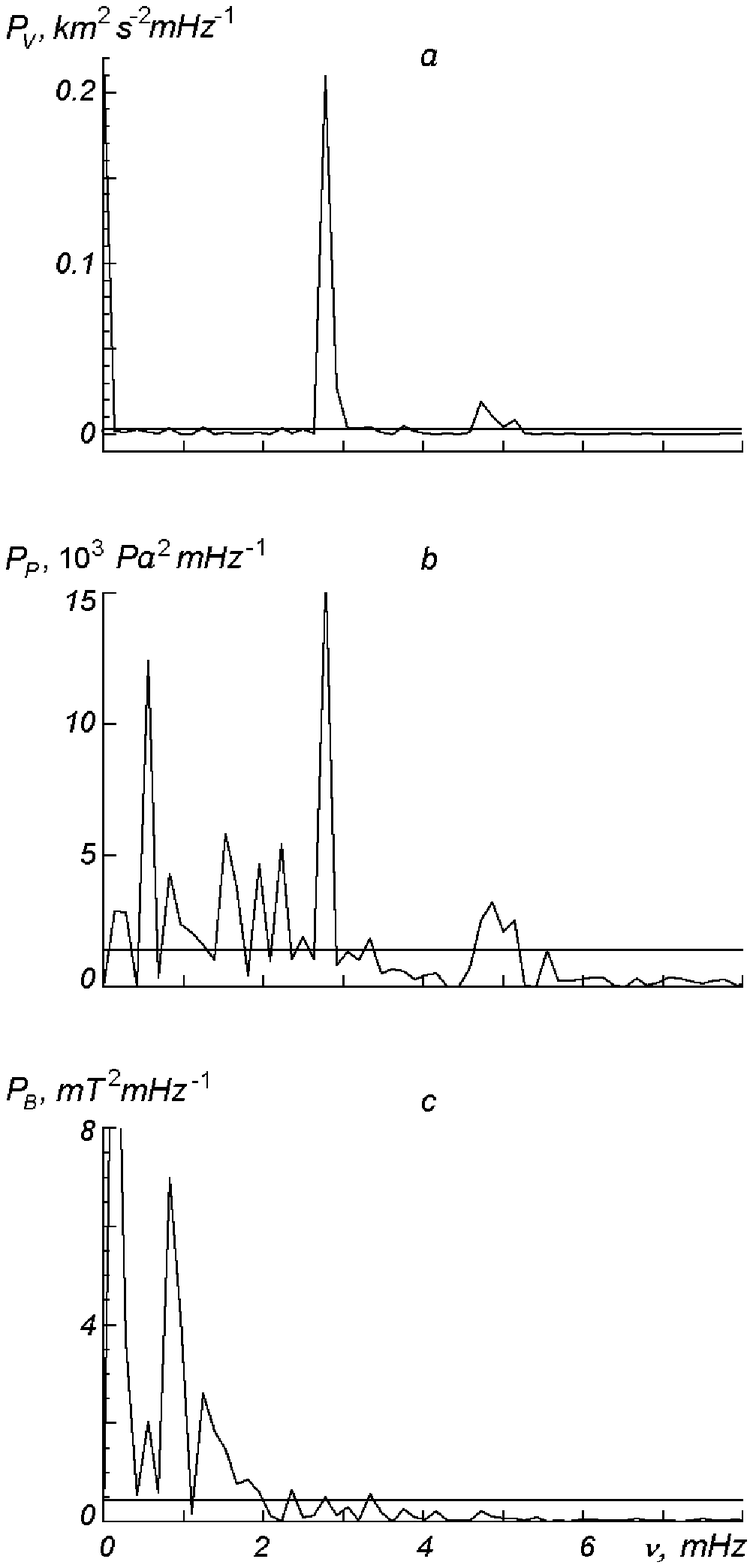}
%}
 \hfill
\vspace{0.1cm}
\parbox[b]{9.cm}{\vspace{0.1cm}
 \caption
{Power spectra of vertical velocity $V_z$ oscillations (a), gas pressure
fluctuations $\delta_P$ (b), and field strength fluctuations
$\delta_B$ at the level $\log\tau_5 = 0$ (c). Straight line)
99-percent confidence level determined in accordance with
\cite{23}.} \label{F-7}}
 \end{figure}
%%%%%%%%%%%%%%%%%%%%%%%%%%%%%%%%%%%%%%%%%%%%%%%%%%%%%%%%%%%%

Velocity oscillations with frequencies of 2.57, 3.88, and 5.58
mHz were also obtained in \cite{43} from a 3-D simulation of the
solar surface convection in the absence of magnetic field. The
p-mode oscillations were assumed to be excited by random
nonadiabatic pressure fluctuations near the Sun's surface. Such
fluctuations are produced by the radiative cooling which can
locally deviate for a short time from equilibrium with the
heating produced by convective motions. Figure 6a shows the gas
pressure fluctuations $\delta P_g$ calculated for our models.
Here $P_g$ is the pressure averaged over one model at a specific
point in time and <$P_g$> is the pressure averaged over all
models in a time interval of 120 min. Comparison of the velocity
and pressure oscillations suggests that there should be a
relationship between them. In the power spectrum calculated for
pressure oscillations (Fig.~7b) there are power peaks at the
same frequencies as in the spectrum of velocity oscillations in
the 5-min and 3-min bands.

To examine the influence of magnetic field on the oscillations
of vertical velocities, we obtained the field strength
fluctuations $\delta B$ (Fig.~6b) from the simulation data and
calculated  their power spectrum (Fig.~7c). First of all we
point out a constant increase of the magnetic field and its
well-marked oscillations associated with the evolution and
disruption of strong flux tubes in the simulation region. A
strong peak in the field strength oscillation power was found at
$\nu = 0.83$~mHz (20 min) and a smaller peak was found at 1.25
mHz (13 min). When comparing the magnetic field oscillations
with vertical velocity oscillations, we found that the negative
velocity component almost disappeared and the positive component
increased when the magnetic field was stronger, that is, the
upflows were slower and downflows faster. This means that the
velocity oscillations are nonlinear. Such quasi-oscillatory
motions in flux tubes were predicted in \cite{21}. Nonlinear
oscillations were considered there to be a superposition of
upflows and downflows with different amplitudes and durations.
The upflow velocities were estimated there at $-0.5$~km~s$^{-1}$
and downflow velocities at about 2~km~s$^{-1}$. Our results for
the moments of magnetic field strengthening are very close to
these data. As seen in Fig.~6a, the magnetic field modulates the
velocity oscillations. Based on the results of MHD simulation
\cite{4}, we attribute this effect to convective instability in
intergranular lanes. About 10--12 min after the formation of a
strong flux tube begins, conditions favorable for convective
collapse set up, and the collapse produces a kilogauss field.
Then the flux tube is in a quasi-stable state for some time
($\leq 10$~min), it is narrower and closer to the vertical, its
temperature grows, the field is stronger, and the Wilson
depression correlates with the vertical velocity. Oscillatory
downward motions become more intense in the flux tube region.
After each abrupt downward shift of magnetic configuration,
there comes a period of a slacker convective collapse. These
oscillations persist until the strong evacuation in the upper
part of the flux tube begins to destroy the tube (i.e., until
the beginning of reverse convective collapse). As this takes
place, downflows are replaced by upflows with supersonic
velocities. The flux tube rapidly dissipates in the course of
several minutes. The size and the inclination of the flux tube
increase, the magnetic field and the temperature diminish, the
Wilson depression also diminishes. As the dynamical condition
changes, the inclination of flux tubes also changes with time
--- it oscillates. This can give rise to waves propagating along
the flux tubes \cite{44}.

So, the convective and superconvective instability processes
going in strong flux tubes bring about a~nonlinearity in the
oscillatory motions of magnetized plasmas in the simulation
region.

%%%%%%%%%%%%%%%%%%%%%%%%%%%%%%%%%%%%%%%%%%% Figure 8
  \begin{figure}
\centerline{\hspace{1.5cm}
\includegraphics [scale=.9]{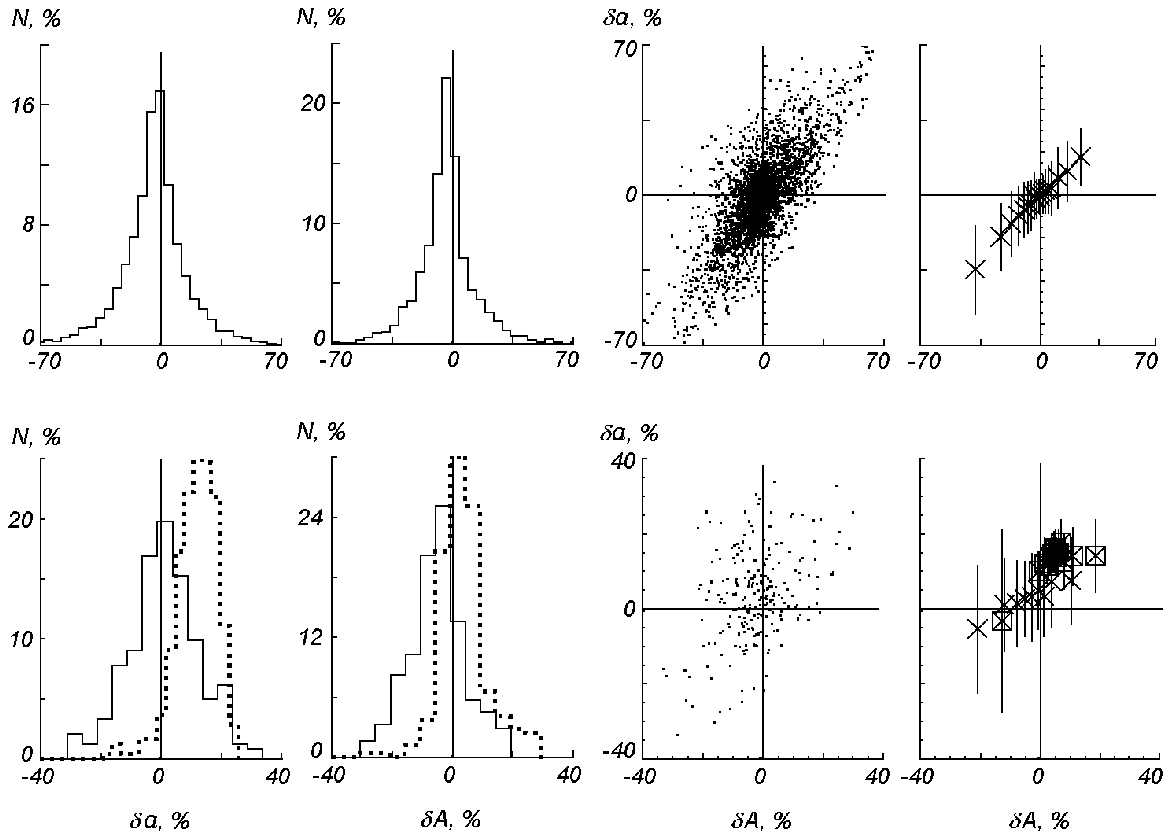}}
 \hfill
%
%\vspace{0.1cm}
 \caption
{Distributions  of amplitude asymmetry $\delta a$ and area
asymmetry $\delta A$ of $V$ profiles, scatter of asymmetry
parameters and their correlations: upper plots) all profiles,
lower plots) profiles averaged with a 700-km step. Dotted line
and square) FTS observation data.} \label{F-8}
 \end{figure}
%%%%%%%%%%%%%%%%%%%%%%%%%%%%%%%%%%%%%%%%%%%%%%%%%%%%%%%%%%%%
{\bf Asymmetry.} Figure 8 demonstrates the distribution of the
amplitude and area asymmetries, $\delta a$ and $\delta A$, and
their correlations. For unaveraged profiles, the amplitude
asymmetry varies in a wide range ($\pm 70$~percent) at small
amplitudes. The area asymmetry has a similar distribution, but
the scatter is smaller. The amplitude asymmetry correlates with
the area asymmetry, at positive values they vary in nearly
direct proportion. This means that both asymmetries are caused
by the same factors -- the velocity field and magnetic field
gradients. Mean asymmetries are close to zero ($\overline{\delta
a}=-1$, $\overline{\delta A}=1$ percent for unaveraged profiles
and $\overline{\delta a}=3$, $\overline{\delta A}=-2$ percent
for averaged ones). Spatial averaging drastically weakens the
correlation between $\delta a$ and $\delta A$ and changes the
inclination of the relation --- the amplitude asymmetry becomes
more positive. This suggests that the spatial averaging is also
a cause of observed profile asymmetry, the amplitude asymmetry
being more sensitive to the averaging. This may be the reason
why the amplitude asymmetry measured from observations is
greater, as a rule, than the area asymmetry: $\delta a = 15$
percent and $\delta A = 6$ and 4 percent for the Fe~I lines
$\lambda\lambda$~630.15 and 630.25~nm \cite{37,20}. Figure 8
also shows the asymmetries found in FTS observations \cite{48},
they are much greater, on the average, than those obtained in
our calculations. This discrepancy seems to be caused by a
substantial difference in the spatial and temporal averaging and
by different activity levels in the observed and simulated
regions.

%%%%%%%%%%%%%%%%%%%%%%%%%%%%%%%%%%%%%%%%%%% Figure 9
  \begin{figure}
\centerline{\hspace{3.cm}
\includegraphics [scale=.85]{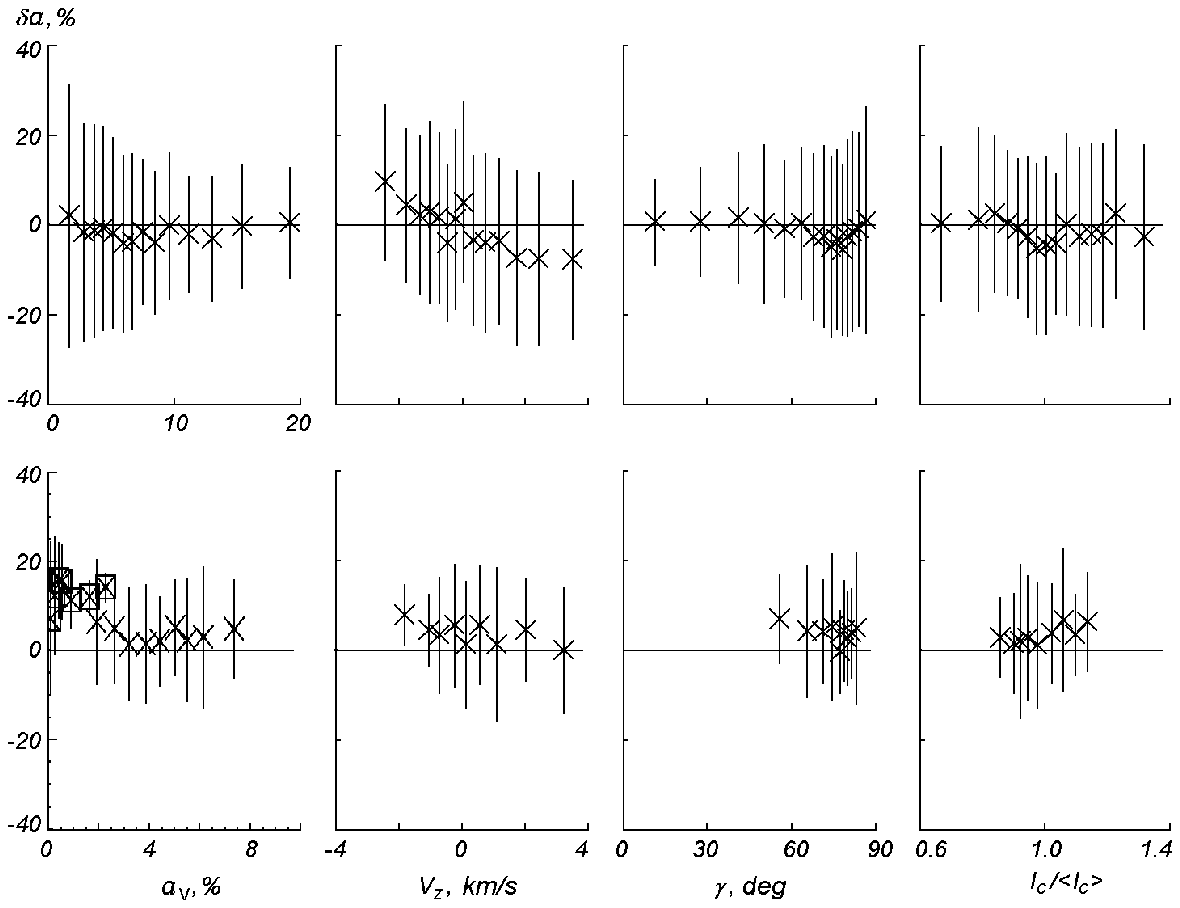}}
 \hfill
\vspace{0.1cm}
 \caption
{Amplitude asymmetry as a function of mean amplitude of $V$
profile, vertical velocity, field inclination angle, and
continuum contrast: upper plots) all profiles, lower plots)
profiles averaged with a 700-km step.} \label{F-9}
 \end{figure}
%%%%%%%%%%%%%%%%%%%%%%%%%%%%%%%%%%%%%%%%%%%%%%%%%%%%%%%%%%%%

The observed area asymmetry is a factor of 2--3 greater than the
amplitude asymmetry. This difference is yet to be explained. Our
results suggest that it is caused by insufficient spatial
resolution of observations and by atmospheric effects. When the
quality of observations and the spatial resolution are upgraded,
the relative number of strong $V$ profiles observed in the
kilogauss flux tube will increase. Our calculations demonstrate
that the asymmetry in such profiles is much smaller because
there are no sudden velocity and field strength gradients inside
flux tubes as compared to their periphery or the regions of weak
turbulence fields of mixed polarity. These properties of strong
$V$ profiles also account for a small scatter in their
asymmetries, while the scatter for weak $V$ profiles can be as
large as 70 percent.

We analyzed the relationship between the $V$-profile asymmetry
and the amplitude $a_V$, velocity $V_z$, inclination angle
$\gamma$, and contrast (Fig.~9). The correlation of $\delta a$
with these parameters is rather weak; the only feature worthy of
notice is an appreciable growth of positive asymmetry with the
velocity of upflows while downflows are more often associated
with negative amplitude asymmetries.

{\bf Center-to-limb variation.} Investigations of the
center-to-limb variations in the $V$-profile shifts and
asymmetries involve a huge amount of calculations, and so we
selected  ten snapshots only. Figure 10 shows the  velocity
$V_z$ as functions of their position on the disk
($\mu=\cos\theta$). $V_z$ is positive (predominance of
downflows) outside the disk center, it amounts up to 2
km~s$^{-1}$, on the average, at $\mu = 0.9$ and changes almost
not at all to the limb.
%%%%%%%%%%%%%%%%%%%%%%%%%%%%%%%%%%%%%%%%%%% Figure 10
  \begin{figure}
\centerline{\hspace{1.5cm}
\includegraphics [scale=0.8]{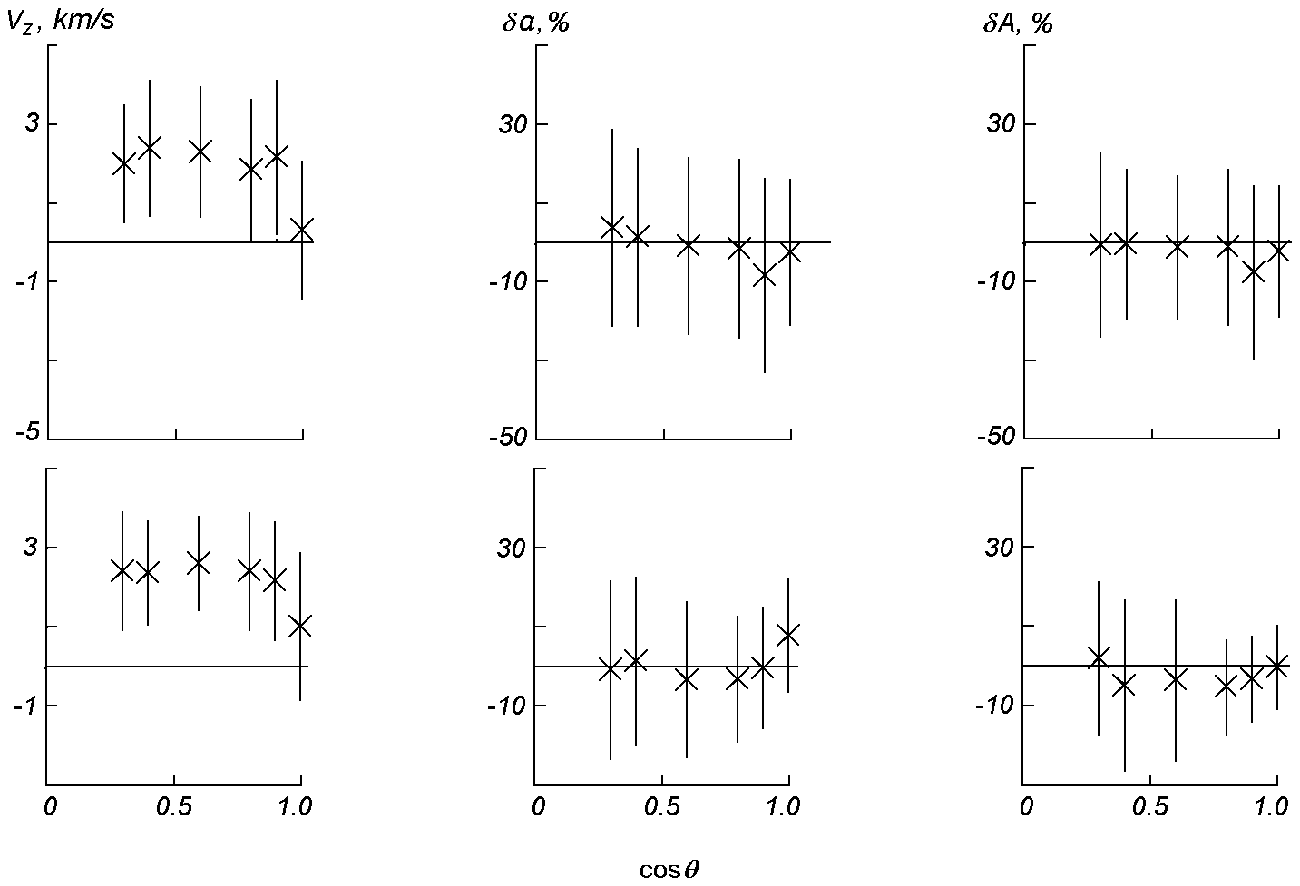}}
 \hfill
\vspace{0.1cm}
 \caption
{Center-to-limb variations of vertical velocity, amplitude
asymmetry, and area asymmetry: upper plots) all profiles, lower
plots) profiles averaged with a 700-km step.} \label{F-10}
 \end{figure}
%%%%%%%%%%%%%%%%%%%%%%%%%%%%%%%%%%%%%%%%%%%%%%%%%%%%%%%%%%%%

Spatial averaging slightly affects the $V_z$--$\mu$ dependence.
According to observations \cite{32}, downflows dominate at the
disk center, and the profile shifts are close to zero at the
limb. Our results can differ from observations for various
reasons. We suppose that the velocity increase obtained in our
calculations is a result of the canopy effect in the upper parts
of flux tubes, where the profiles calculated for $\cos\theta
\approx 1$ are formed. According to the 2D MHD simulations used
in this paper, the strong flows of matter in the upper layers
are predominantly inclined, they are directed slightly
downwards, to the centers of intergranular lanes (see Fig.~1).
Another reason for the discrepancy can be related to some
peculiarities of the 2D MHD models -- the upper boundary
conditions imposed on the upper layers in the models do not
correspond to actual conditions in the upper photosphere. This
may affect the calculations of the line profiles far from the
disk center, these profiles being formed higher in the
photosphere than those which are located closer to the disk
center. At the same time the accuracy of observations
deteriorates closer to the limb. It should also be noted that
there is observational evidence of high horizontal velocities
(up to 2~km~s$^{-1}$) \cite{51}, and they are in better accord
with our results.

The center-to-limb variations of mean asymmetry $\delta a$,
particularly the asymmetry for averaged profiles, agree
satisfactorily with observation data. In FTS observations
\cite{50} the amplitude asymmetry diminished from 10--15 percent
at the disk center to zero at the limb. In \cite{20} the mean
asymmetry $\delta a$ varied from 15 percent (center) to $-5$
percent (limb), with the intersection point near $\cos\theta =
0.7$. Similar results were obtained in \cite{32}, but $\delta a$
and $\delta A$ slightly increased at $\cos\theta = 0.86$ and
then monotonically diminished to the limb.

The area asymmetry in our calculations is small and
predominantly negative (Fig.~10), while in the FTS observations
in \cite{50} the asymmetry is positive at the center, diminishes
to the limb and becomes negative. Negative area asymmetries were
observed in \cite{32} at the disk center in regions with small
filling factor.

One can see in Fig.~10 that the spatial averaging of profiles
affected the most the amplitude asymmetry, especially at the
disk center. We wish to stress once more that the amplitude
asymmetry is the parameter most sensitive to spatial averaging.

\section{Conclusion}
     \label{Conclusion}
We used the results of  two-dimensional MHD simulation of solar
granulation to investigate  the motions of matter in magnetic
elements  with high spatial resolution (35~km).
The analysis of the $V$ profiles of the Fe~I  $\lambda$
630.25~nm line synthesized in snapshots of the MHD simulations
allowed us to make some inferences as to the vertical velocity
of magnetized plasma motions in photospheric regions with a mean
magnetic flux density of 0.2~mT and mean unsigned field strength
of 35~mT.

1. The mean velocity is $0.5\pm 2$~km~s$^{-1}$ with a scatter
from $-3$ to 9~km~s$^{-1}$. Downflows with a mean velocity of
$3\pm 2$~km~s$^{-1}$ dominate in intergranular regions inside
strong vertical flux tubes, while upflows with a mean velocity
of $0.5\pm 2$~km~s$^{-1}$ are typical of granular regions
outside flux tubes.

2. There is a noticeable correlation between the velocity and
the amplitude and area asymmetries of $V$ profiles. The positive
asymmetry increases, on the average, with upflow velocity. The
negative asymmetry occurs most often in the profiles formed at
the sites with dominant downflows. The mean asymmetry of
amplitudes and areas is about 1 percent with a scatter of 70
percent in weak $V$ profiles and 10 percent in strong ones.

3. The mean velocity fluctuates nonlinearly in time. There are
two peaks in the velocity power spectrum --- a stronger peak in
the 5-min band and a much weaker one in the 3-min band. Velocity
fluctuations correlate with gas pressure fluctuations at the
surface level, and they also depend on magnetic field
fluctuations. The power spectrum of field strength fluctuations
has two peaks at periods of 20 and 13 min. Periodic variations
of magnetic field strength are related to intensification and
dissipation of magnetic fields in strong flux tubes. When the
field strength increases, oscillatory downward motions are
intensified, and this results in a nonlinearity of mean velocity
oscillations.

4. The mean velocity substantially changes when going from the
solar disk center to the limb. These changes occur due to
intense slopping motions in the upper layers of simulation
region. The mean velocity increases to 2~km~s$^{-1}$ at a small
distance from the disk center ($\cos\theta = 0.9$), and then it
slightly changes to the limb. The amplitude asymmetry decreases
and changes its sign from positive to negative when going from
the center to the limb of the solar disk. The negative area
asymmetry is greater at the limb.

5. The mean velocity depends on spatial averaging of profiles.
At lower resolutions the $V$-profile amplitudes and other
parameters substantially change, and they are closer to the
observed parameters. The asymmetry of $V$ profiles is the
parameter most sensitive to spatial averaging, the amplitude
asymmetry growing much faster at lower resolutions than the area
asymmetry. This may be a reason why observed amplitude
asymmetries are greater than area asymmetries.

\vspace{1.cm}
 {\bf Acknowledgements.}
We wish to thank S. Solanki for the possibility to work at the
Max Planck Institute for Aeronomy, for making available FTS
observation data, and for useful comments, S. Ploner for his
thorough support rendered to the author during her stay at the
Max Planck Institute for Aeronomy and for useful discussion, M.
Sch\"{u}ssler for discussion of the results, and N. V.
Kharchenko for her useful advice on the statistics techniques.
We also wish to thank the referee for useful critical comments.
The research described in this publication is a part of an
international cooperation program, it was made possible in part
by Grant No. CLG97501 from NATO and Grant No. 00084 from INTAS.

%========-------------------------------------------------

%%%%%%%%%%%%%%%%%%%%%%%%%%%%%%%%%%%%%%%%%%%%%%%%%%%%%%%%%%%%

 %%%%%%%%%%%%%%%%%%%%%%%%%%%%%%%%%%%%%%


\begin{thebibliography}{99}


\bibitem{15}
M. A. Amer and F. Kneer, ``High spatial resolution
spectropolarimetry of small-scale magnetic elements on the
Sun,'' Astron. and Astrophys., vol. 273, no. 1, pp. 304--312,
1993.

\bibitem{1} I. N. Atroshchenko and V. A. Sheminova, ``Numerical simulation
of the interaction between solar granules and small-scale
magnetic fields,'' Kinematika i Fizika Nebes. Tel [Kinematics
and Physics of Celestial Bodies], vol. 12, no. 4, pp. 32--45,
1996.

\bibitem{2}
I. N. Atroshchenko and V. A. Sheminova, ``Simulation of spectral
effects with the use of two-dimensional magnetohydrodynamic
models of the solar photosphere,'' Kinematika i Fizika Nebes.
Tel [Kinematics and Physics of Celestial Bodies], no. 5, pp.
32--47, 1996.


\bibitem{16}
T. E. Berger, C. A. Schrijver, R. A. Shine, et al., ``New
observations of subarcsecond photospheric bright points,''
Astrophys. J., vol. 454, no. 1, pp. 531--544, 1995.

\bibitem{3}
P. N. Brandt, A. S. Gadun, and V. A. Sheminova, ``Absolute
shifts of Fe~I and Fe~II lines in solar active regions (disk
center),'' Kinematika i Fizika Nebes. Tel [Kinematics and
Physics of Celestial Bodies], vol. 13, no.5, pp. 75--86, 1997.

\bibitem{4} A. S. Gadun,
``Two-dimensional nonstationary magnetogranulation,'' Kinematika
i Fizika Nebes. Tel [Kinematics and Physics of Celestial
Bodies], vol. 16, no. 2, pp.~99--120, 2000.

\bibitem{5} A. S. Gadun, V. A. Sheminova, and S. K. Solanki,
``Formation of small-scale magnetic elements: surface
mechanism,''Kinematika i Fizika Nebes. Tel [Kinematics and
Physics of Celestial Bodies], vol. 15, no. 5, pp. 387--397,
1999.

\bibitem{17}
A. S. Gadun, S. K. Solanki, and A. Johannesson, ``Granulation
near the solar limb: observations and 2D modelling,'' Astron.
and Astrophys., vol. 350, no. 3, pp. 1018--1034, 1999.



\bibitem{18}
A. S. Gadun, S. K. Solanki, V. A. Sheminova, and S. R. O.
Ploner, ``A formation mechanism of magnetic elements in regions
of mixed polarity,'' Solar Phys., vol. 203, no. 1, pp. 1--7,
2001.



\bibitem{19}
S. I. Gopasyuk, V. A. Kotov, A. B. Severny, and T. T. Tsap,
``The comparison of the magnetographic magnetic field measured
in different spectral lines,'' Solar Phys., vol. 31, no. 2, pp.
307--316, 1973.


\bibitem{20} U. Grossmann-Doerth, C. U. Keller, and M. Sch\"{u}ssler,
``Observations of the quiet Sun's magnetic field,'' Astron. and
Astrophys., vol. 315, no. 3, pp. 610--617, 1996.


\bibitem{21} U. Grossmann-Doerth, M. Sch\"{u}ssler, and S. K.
Solanki, ``The effect of non-linear oscillations in magnetic
flux tubes of Stokes V asymmetry,'' Astron. and Astrophys., vol.
249, no. 1, pp. 239--242, 1991.

\bibitem{22}U. Grossmann-Doerth, M. Sch\"{u}ssler, and O. Steiner,
``Convective intensification of solar surface magnetic fields:
Results of numerical experiments,'' Astron. and Astrophys., vol.
337, no. 3, pp. 928--939, 1998.


\bibitem{23}
E. J. Groth, ``Probability distribution related to power
spectra,'' Astrophys. J., Suppl. Ser., vol.~286, no. 1, pp.
285--302, 1975.


\bibitem{6}
V. A. Kotov, N. N. Stepanyan, and Z. F. Shcherbakova, ``Role of
the background magnetic field and the fields of active regions
and spots in the general magnetic field of the Sun,'' Izv. Krym.
Astrofiz. Observ., vol. 56, pp. 75--83, 1977.


\bibitem{24}
R. Howard and J. O. Stenflo, ``On the filamentary nature of
solar magnetic fields,'' Solar Phys., vol. 22, no. 2, pp.
402--417, 1972.



\bibitem{25}
C. U. Keller, S. K. Solanki, T. D. Tarbel, et al., ``Solar
magnetic field strength determinations from high spatial
resolution filtergrams,'' Astron. and Astrophys., vol. 236, no.
1, pp. 250--255, 1990.



\bibitem{26}
C. U. Keller, F.-L. Deubner, U. Egger, et al., ``On the strength
of solar intra-network fields,'' Astron. and Astrophys., vol.
286, no. 2, pp. 626--634, 1994.



\bibitem{27}
E. V. Khomenko, M. Collados, S. K. Solanki, et al., ``Quiet-Sun
intra-network magnetic fields observed in the infrared,''
Astron. and Astrophys., vol. 408, no. 3, pp. 1115--1135, 2003.



\bibitem{28}
K. D. Leka and O. Steiner, ``Understanding small solar magnetic
structures: comparing numerical simulations to observation
fields,'' Astrophys. J., vol. 552, no. 1, pp. 354--371, 2001.



\bibitem{29}
H. Lin and T. Rimmele, ``The granular magnetic fields of the
quiet Sun,'' Astrophys. J., vol. 514, no. 1, pp.~448--455, 1999.


\bibitem{30}
B. W. Lites, ``Characterization of magnetic flux in the quiet
Sun,'' Astrophys. J., vol. 573, no. 1, pp.~31--444, 2002.



\bibitem{31}
B. W. Lites, A. Skumanich, and V. Martinez Pillet, ``Vector
magnetic fields of emerging solar flux. I. Properties at the
solar site of emergence,'' Astron. and Astrophys., vol. 333, no.
3, pp. 1053--1068, 1998.



\bibitem{32}
V. Martinez Pillet, B. W. Lites, and A. Skumanich, ``Active
region magnetic fields. I. Plage fields,'' Astrophys. J., vol.
474, no. 2, pp. 810--842, 1997.


\bibitem{33} S. R. O. Ploner, M. Sch\"{u}ssler, S. K. Solanki,
and A. S. Gadun, ``An example of reconnection and magnetic flux
recycling near the solar surface,'' in: Theory, Observation, and
Instrumentation, M. Sigwarth (Editor), ASP Conf. Ser., vol. 236,
pp. 363--370, 2001.


\bibitem{34} S. R. O. Ploner, M. Sch\"{u}ssler, S. K. Solanki
et al., ``The formation of one-lobe Stokes $V$ profiles in an
inhomogeneous atmosphere,'' ibid., pp. 371--378, 2001.


\bibitem{35}
S. R. O. Ploner, S. K. Solanki, and A. S. Gadun, ``The evolution
of solar granules deduced from 2D simulations,'' Astron. and
Astrophys., vol. 352, no. 2, pp. 679--696, 1999.

\bibitem{7}
D. N. Rachkovskii and T. T. Tsap, ``Study of magnetic fields by
the method of the ratio of field strengths in lines measured
outside active regions on the Sun,'' ibid., vol. 71, pp. 79--87,
1985.


\bibitem{36}
M. Sch\"{u}ssler, ``Magneto-convection,'' in: Magnetic Fields
Across the Hertzsprung-Russell Diagram, G. Mathys, S. K.
Solanki, and D. T. Wickramasinghe (Editors), ASP Conf. Ser.,
vol. 258, pp.~115--123, 2001.

\bibitem{8}
A. B. Severnyi, ``On the nature of magnetic fields on the Sun
(fine field structure),'' Astron. Zhurn., vol. 42, no. 2, pp.
217--232, 1965.

\bibitem{10}
V. A. Sheminova, Calculating Stokes Profile Parameters of
Magnetic Absorption Lines in Stellar Atmospheres [in Russian],
Kyiv, 1990 (VINITI File No. 2940-B90, 30 May 1990).


\bibitem{11}
V. A. Sheminova, ``Two-dimensional MHD models of solar
magnetogranulation. Testing of models and methods of Stokes
diagnostics,'' Kinematika i Fizika Nebes. Tel [Kinematics and
Physics of Celestial Bodies], vol. 15, no. 5, pp. 398--412,
1999.


\bibitem{12}
V. A. Sheminova and A. S. Gadun, ``Convective shifts of iron
lines in the spectrum of the solar photosphere,'' Kinematika i
Fizika Nebes. Tel [Kinematics and Physics of Celestial Bodies],
vol. 18, no. 1, pp. 18--32, 2002.


\bibitem{13}
V. A. Sheminova, ``The Fe~I $\lambda$ 1564.8~nm line and the
distribution of solar magnetic fields,'' Kinematika i Fizika
Nebes. Tel [Kinematics and Physics of Celestial Bodies], vol.
19, no. 2, pp. 107--125, 2003.


\bibitem{14}
V. A. Sheminova and A. S. Gadun, ``Evolution of solar magnetic
flux tubes from Stokes parameter observations,'' Astron. Zhurn.,
vol. 77, no. 10, pp. 790--800, 2000.


\bibitem{37} M. Sigwarth, K. S. Balasubramaniam, M. Kn\"{o}lker,
and W. Schmidt, ``Dynamics of solar magnetic elements,'' Astron.
and Astrophys., vol. 349, no. 3, pp. 941--955, 1999.


\bibitem{38}
S. K. Solanki, ``Stokes V asymmetry and shift of spectral
lines,'' Astron. and Astrophys., vol. 221, no. 2, pp.~338--341,
1989.



\bibitem{39}
S. K. Solanki, ``Small-scale solar magnetic fields: an
overview,'' Space Sci. Rev., vol. 31, pp.~1--188, 1993.



\bibitem{40}
S. K. Solanki, ``Small-scale photospheric structure of the solar
magnetic fields outside sunspots,'' in: Magnetic Fields Across
the Hertzsprung-Russell Diagram, S. Mathys, S. K. Solanki, and
D. T. Wick- ramasinghe (Editors), ASP Conf. Ser., vol. 258, pp.
45--53, 2001.


\bibitem{41}
S. K. Solanki and J. O. Stenflo, ``Velocities in solar magnetic
flux tubes,'' Astron. and Astrophys., vol. 170, no. 1--2, pp.
311--329, 1986.



\bibitem{42}
R. F. Stein and~\AA. Nordlund, ``Simulation of solar
granulation. I. General properties,'' Astrophys. J., vol. 499,
no. 2, pp. 914--933, 1998.



\bibitem{43}
R. F. Stein and~\AA. Nordlund, ``Solar oscillations and
convection. II. Excitation of radial oscillations,'' Astrophys.
J., vol. 546, no. 1, pp. 585--603, 2001.



\bibitem{44} O. Steiner, U. Grossmann-Doerth, M. Kn\"{o}lker,
and M. Sch\"{u}ssler, ``Dynamical interaction of solar magnetic
elements and granular convection: results of a numerical
simulation,''  Astrophys. J., vol. 495, no. 1, pp.~468--484,
1998.

\bibitem{45}
J. O. Stenflo, ``Magnetic-field structure of the photospheric
network,'' Solar Phys., vol. 32, no. 1, pp. 41--63, 1973.

\bibitem{46}
J. A. Stenflo, ``A model of the supergranulation network and of
active-region plages,'' Solar Phys., vol. 4, no. 1, pp. 79--105,
1975.

\bibitem{47}
J. O. Stenflo and J. W. Harvey, ``Dependence of the properties
of magnetic fluxtubes on area factor or amount of flux,''  Solar
Phys., vol. 95, no. 1, pp. 99--118, 1985.

\bibitem{48} J. O. Stenflo, J. W. Harvey, J. W. Brault, and S. K.
Solanki, ``Diagnostics of solar magnetic fluxtubes using a
Fourier transform spectrometer,'' Astron. and Astrophys., vol.
131, no. 2, pp. 333--346, 1984.

\bibitem{49} J. O. Stenflo, C. U. Keller, and A. Candorfer,
``Differential Hanle effect and the spatial variation of
turbulent magnetic fields on the Sun,'' ibid., vol. 329, no. 1,
pp. 319--328, 1998.

\bibitem{50} J. O. Stenflo, S. K. Solanki, and J. W. Harvey,
``Center-to-limb variation of Stokes profiles and the
diagnostics of solar magnetic fluxtubes,'' Astron. and
Astrophys., vol. 171, no. 2, pp. 305--316, 1987.

\bibitem{9}
T. T. Tsap and I. S. Laba, ``Magnetic fields and vertical
motions in supergranules,'' Izv. Krym. Astrofiz. Observ., vol.
73, pp. 62--70, 1985.

\bibitem{51} R. Volmer, F. Kneer, and C. Bendlin, ``Short period waves
in small-scale magnetic flux tubes on the Sun,''Astron. and
Astrophys., vol. 34, no. 1, pp. L1--L4, 1995.

\bibitem{52} H. Zirin and R. Cameron, ``Properties of the quiet-Sun
magnetic fields as revealed by spectrovideomagnetograph,'' Il
Nuovo Cimento, vol. 25C, no. 5--6, pp. 557--563, 2002.



\end{thebibliography}
\end{document}